\documentclass[%
floatfix,
 aip,
 pof,
 amsmath,amssymb,
 preprint,
]{revtex4-1}
\usepackage{mathrsfs}
\usepackage{dcolumn}
\usepackage{bm}
\usepackage[dvipsnames]{xcolor}
\usepackage{mathptmx}
\usepackage{etoolbox}
\usepackage{siunitx}
\usepackage{caption}
\usepackage{graphicx}
\usepackage{bm}
\usepackage[utf8]{inputenc}
\usepackage[T1]{fontenc}
\usepackage{etoolbox}
\usepackage{amsmath}
\usepackage{xcolor}
\usepackage{epsfig}
\usepackage{epstopdf}
\usepackage{multirow}
\usepackage{booktabs}

\usepackage{ragged2e} 

\usepackage{amssymb}

\begin{document}

\title[]{Density evolution at fluid--fluid interfaces from a generalized Gibbs--Duhem relation}

\author{Fei Wang}

\affiliation{Institute of Nanotechnology, 
Karlsruhe Institute of Technology, 
Hermann-von-Helmholtz-Platz 1,
76344 Eggenstein-Leopoldshafen, Germany}

\email{fei.wang@kit.edu; ORCID: 0000-0003-3318-3264}

\date{\today}

\begin{abstract}
The classical Gibbs--Duhem relation is restricted to quasistatic thermodynamic processes and does not include kinetic energy contributions. Consequently, its connection to Newtonian fluid mechanics remains indirect. We derive a generalized Gibbs--Duhem relation that consistently incorporates kinetic effects, thereby linking thermodynamic and mechanical descriptions of fluid motion. On this basis, we formulate a density evolution equation for fluid--fluid interfaces. The proposed equation recovers, as limiting cases, the classical relations for the speed of sound, Bernoulli's equation, and the van der Waals equation of state.
\end{abstract}

\maketitle

\section{Introduction} 

Classical continuum mechanics describes fluid motion through Newton's second law~\cite{landau2013fluid},
\begin{equation}
\rho \frac{\mathrm d \mathbf {u}}{\mathrm d t}=-\nabla p,
\label{eq:1}
\end{equation}
where $\rho$ is the mass density, $\mathbf{u}$ is the velocity field, and $p$ is the pressure. This formulation has been remarkably successful for a wide range of continuum flows. However, fluid--fluid, solid--fluid, and solid--solid interfaces involve strong coupling between mechanics, mass transport, and thermodynamics, motivating formulations that consistently incorporate these effects.

A central thermodynamic relation is the Gibbs--Duhem equation~\cite{sprik2025thermodynamics},
\begin{equation}
\mathrm d p+\phi \mathrm d\mu+s_\tau\mathrm d\tau=0,
\end{equation}
where $\mu$ is the chemical potential, $\phi$ is the volume fraction, $\tau$ is the absolute temperature, and $s_\tau$ is the entropy density. The classical Gibbs--Duhem relation is derived for thermodynamic equilibrium, or quasistatic processes, and therefore does not explicitly include the kinetic-energy contribution associated with fluid motion. Consequently, its connection with Newtonian fluid mechanics is indirect. In this work, we derive a generalized Gibbs--Duhem relation that incorporates kinetic energy, thereby establishing a thermodynamic framework that is directly compatible with fluid motion.

A second issue concerns the evolution of mass density. Equation~\eqref{eq:1} governs the acceleration of a material element but does not itself provide an evolution equation for the density. In many continuum formulations, this issue is circumvented by adopting the incompressibility constraint~\cite{yue2004diffuse,worner2012numerical},
\begin{equation}
\frac{\mathrm d\rho}{\mathrm dt}=0,
\end{equation}
which is well justified for many single-phase liquids and solids. At fluid--fluid interfaces, however, diffusion, mixing, and excess-volume effects generally produce local density variations, making an explicit density-evolution equation desirable.

A third consideration concerns the kinematic description of transport. Conventional continuum mechanics is formulated primarily in the Eulerian framework, in which the conservation laws are expressed as
\begin{align}
\frac{\mathrm D\rho}{\mathrm Dt}
&=
\frac{\partial\rho}{\partial t}
+\nabla\cdot(\rho\mathbf u),\\
\frac{\mathrm D(\rho\mathbf u)}{\mathrm Dt}
&=
\frac{\partial(\rho\mathbf u)}{\partial t}
+\nabla\cdot(\rho\mathbf u\otimes\mathbf u).
\end{align}
These equations are obtained by applying Gauss's theorem to an arbitrary control volume and provide an exact mathematical representation of mass and momentum conservation. In the present work, we instead adopt a Lagrangian description, in which material elements are tracked directly. This formulation is particularly convenient for coupling density evolution with the thermodynamic description developed below.

Motivated by these considerations, we derive a generalized Gibbs--Duhem relation that incorporates kinetic-energy contributions and formulate a corresponding density-evolution equation for fluid--fluid interfaces. The resulting framework provides a thermodynamically consistent description of the coupled evolution of density, momentum, and interfacial transport. In appropriate limiting cases, the theory recovers the classical expressions for the speed of sound, Bernoulli's equation, and the van der Waals equation of state, thereby demonstrating consistency with established fluid mechanics and thermodynamics.

\section{Setup and concept}

\begin{figure}[h!]
 \centering
 \includegraphics[width=0.8\linewidth]{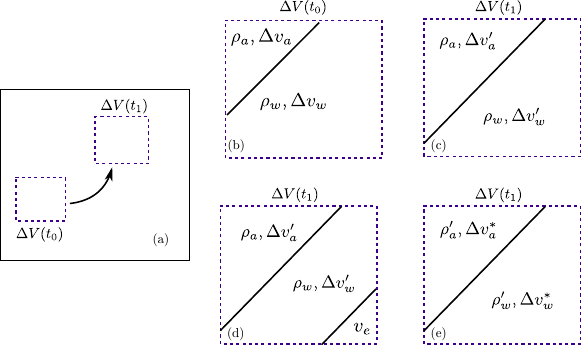}
 \caption{
 (a) The transport of the volume element $\Delta V$ in the Lagrange framework.
 (b) At time $t_0$, 
 the material volume element $\Delta V$ consists of an air volume 
$\Delta v_a$.
and a water volume $\Delta v_w$.
 (c)  At time $t_1$,
 the same material volume element $\Delta V$  consists of the updated air 
volume  $\Delta v_a^\prime$
and water volume $\Delta v_w^\prime$.
The volumes $\Delta v_a^\prime$
 and $\Delta v_w^\prime$  include the mass transported into $\Delta V$  through diffusion during the time interval $[t_0, t_1]$.
 (d) The volume element $\Delta V$ contains excess volume $v_e$ instead of the mechanical mixing in (c).
 (e) After redistributing the excess volume $v_e$
 into $\Delta v_w^\prime$ and $\Delta v_a^\prime$, we have the updated partial volumes 
  $\Delta v_w^*$ and $\Delta v_a^*$.
 }
 \label{fig-1}
\end{figure}

We consider a closed system composed of two immiscible fluid phases occupying a domain $V$.
Each phase contains $K \in \mathbb{Z}$ chemical species. 
The interface evolves with a local velocity $\mathbf{u}$ including the mean thermal motion velocity.
 To illustrate the proposed framework, we take the water–air system as an example. 
 The bulk densities of the water and gas phases before mixing are denoted by $\rho_w$ and $\rho_a$, respectively.
 Consistent with van der Waals theory, the water--air interface is assumed to possess a finite thickness within which water and air molecules coexist.
 A volume element $\Delta V$ at time $t_0$ 
 is considered (Fig.~\ref{fig-1}a), containing a statistically sufficient number of fluid particles in the interfacial region to evaluate the local fluid density as (Fig.~\ref{fig-1}b),
\begin{equation}
 \rho=\frac{m_w+m_a}{\Delta V}=\frac{\rho_w \Delta v_w
 +\rho_a \Delta v_a}{\Delta V},
\end{equation}
where $\Delta v_w$ and $\Delta v_a$ denote the partial volumes of the water and gas components, respectively  and $m_w$ and $m_a$ represent the masses of water and gas contained within the volume element.

A common assumption in existing models is that the mixture is locally incompressible, such that (see Fig.~\ref{fig-1}b)
\begin{equation}
 \Delta V =\Delta v_w + \Delta v_a.
\end{equation}
 Under this assumption, the local density $\rho$ can be expressed as a linear interpolation of the volume fractions~\cite{wu2024evolution}
 \begin{equation}
     \rho=\rho_a \phi_a+\rho_w\phi_w,  \phi_i = \Delta v_i /\Delta V,  i=w, a.
     \label{eq:interpolation}
 \end{equation}
 The condition  $\mathrm d \rho/\mathrm d t=0$ 
 is widely adopted in the literature and forms the basis of the incompressible-fluid approximation. Its validity, however, relies on the absence of mass exchange across the boundary of the volume element, e.g., diffusion, as well as overlooking excess volume $v_e$, as discussed below.

When diffusion is present, mass continuously enters and leaves the volume element $\Delta V$, violating the assumption of incompressible fluids based on the density evolution. 
To account for the diffusion effects from time $t_0$
to $t_1$, we rewrite the local volume element as (Fig.~\ref{fig-1}c) 
\begin{equation}
\Delta V=\Delta v_{w}^\prime+\Delta v_{a}^\prime.
\end{equation}
The volume elements $\Delta v_{w}^\prime$ and $\Delta v_{a}^\prime$ 
reflect the mass gained or lost through diffusion within the domain  $\Delta V$.
The associated volume composition is defined as
\begin{equation}
 \phi_w=\frac{\Delta v_{w}^\prime}{\Delta V}\quad \text{and} \quad  \phi_a=\frac{\Delta v_{a}^\prime}{\Delta V}.
 \label{eq:10-0}
\end{equation}
Therefore, the fluid density is calculated as
\begin{equation}
 \rho=\rho_{w}\frac{\Delta v_{w}^\prime}{\Delta V} +\rho_{a}\frac{\Delta v_{a}^\prime}{\Delta V}=\rho_{w}\phi +\rho_{a}(1-\phi).
\label{eq:30-1}
\end{equation}
According  to Eq.~\eqref{eq:30-1},
when diffusion takes place, i.e., $\mathrm d \phi/\mathrm d t\neq0$ (or $\mathrm d \Delta v_{i}^\prime/\mathrm d t\neq0$),
we have $\mathrm d \rho/\mathrm d t\neq0$, when $\rho_w\neq \rho_a$.
Hence, the incompressible condition should be modified by considering the diffusion effect.

In addition, the linear interpolation according to Eq.~\eqref{eq:30-1} 
indicates a zero excess volume of mixing
at every position of the considered domain $V$. 
This type of idealized mechanical mixing contradicts our everyday observations. 
For example, mixing one liter of water with one liter of oil typically does not produce exactly two liters of mixture because of excess volume effects~\cite{reiplinger2022density,caqueret2023density}. Similarly, mixing one liter of water with one liter of gas does not necessarily result in a two-liter water–gas mixture. When considering excess volume $v_e$, 
the volume element is expressed as (Fig.~\ref{fig-1}d)
 \begin{equation}
 \Delta V =\Delta v_{w}^\prime + \Delta v_{w}^\prime+v_e.
\end{equation}
By distributing the excess volume $v_e$
into $\Delta v_w^\prime$ and $\Delta v_a^\prime$,
we have (Fig.~\ref{fig-1}e)
 \begin{equation}
 \Delta V =\Delta v_{w}^* + \Delta v_{a}^*.
\end{equation} 
We define the volume composition including the information of excess volume as
\begin{equation}
 \phi_w=\frac{\Delta  v_w^*}{\Delta  V}\quad \text{and} \quad  \phi_a=\frac{\Delta  v_a^*}{\Delta  V}.
 \label{eq:10-0}
\end{equation}
Therefore, the local fluid density 
is calculated as
\begin{equation}
 \rho=\rho_{w}^\prime\frac{\Delta  v_w^*}{\Delta  V} +\rho_{a}^\prime\frac{\Delta  v_a^*}{\Delta  V}=\rho_{w}^\prime\phi +\rho_{a}^\prime(1-\phi).
\label{eq:30-222}
\end{equation}
Note that both $\rho$ and $\phi$ encode the information associated with the excess volume.
The partial densities $\rho_{w}^\prime$ and $\rho_{a}^\prime$ are implicitly influenced by the excess volume and are therefore no longer constant, 
as they vary with the redistribution of the excess volume. 
Consequently, to fully characterize the state of the system, two evolution equations are required: one governing the volume composition $\phi$, and the other governing the  density $\rho$.

The above concept can be readily extended to the more general case of a multicomponent system, for which the following equality holds:
\begin{equation}
 \phi_i=\frac{\Delta v_i^*}{\Delta V}, \quad \text{and} \quad\sum_i \Delta v_i^*=\Delta V,\quad \text{and} \quad\sum_{i=1}^K \phi_i=1.
\end{equation}
The boundary of the domain $V$ is denoted by $\Gamma$.
For a binary system, we use the notation $\phi_1=\phi$, 
$\phi_2=1-\phi$.
The density of the mixture at any position $\boldsymbol r\in V$
is denoted by $\rho$.

In $n\in \mathbb{Z}$ dimensions, we define the space vector  as 
\begin{equation}
    \bold r(t)=(x_1(t), x_2(t),\cdots, x_n(t)).
\end{equation}
The displacement is expressed as
\begin{equation}
   (\mathrm{d}\bold r)^2 =\sum_{k=1}^n (\mathrm{d} x_k)^2.
   \label{eq:16-9}
\end{equation}

In the following derivation, we use the Lagrangian description 
for the scalar field: $\phi(\bold r(t),t)$
and for the vector field: $\bold u(\bold r(t),t)$.
In this context, the definition of the convection velocity is 
\begin{equation}
    \bold u=\frac{\mathrm d \bold r} {\mathrm d t}.
    \label{eq:10-9}
\end{equation}

The total time derivative is defined as 
\begin{align}
    \frac{\mathrm d \phi }{\mathrm d t}&=\frac{\partial \phi}{\partial t}+\bold u\cdot \nabla\phi;\\
       \frac{\mathrm d \bold u }{\mathrm d t}&=\frac{\partial \bold u}{\partial t}+\bold u\cdot \nabla\bold u.
\end{align}

These definitions differ from the
Eulerian description
for the scale field:
 $\phi(\bold r,t)$
and for the vector field: $\bold u(\bold r,t)$, based on Gauss's theorem.

Moreover, we further emphasize the definition of incompressibility in the Lagrangian framework. In the absence of mass exchange, an incompressible fluid is characterized by the invariance of the volume of each transported material element,
\begin{equation}
\frac{\mathrm{d}(\Delta V)}{\mathrm{d}t}=0.
\end{equation}
Because the mass of each material element is conserved, this condition is equivalent to
\begin{equation}
    \frac{\mathrm{d} \rho}{\mathrm{d}t}=0.
    \end{equation}
When mass exchange between neighboring material elements is allowed, however, these two conditions are no longer equivalent: a material element may retain its volume while its density changes due to mass transfer, such as diffusion. This distinction motivates a revised definition of incompressibility in the presence of mass diffusion.

\section{Theory}

We write the free energy functional of the system as 
\begin{align}
\notag
 E= \int_V e \delta V+\int_\Gamma e_w \delta \Gamma = \int_V [(\theta-\tau s)+ \rho \mathbf{u}\cdot \mathbf{u}]\delta V+
 \int_\Gamma e_w  \delta \Gamma.
\end{align}
In the volume integration, the first part is the Helmholtz-type free energy 
density $f=\theta-\tau s$ 
($\theta$: internal energy excluding thermal kinetic energy, $\tau$: absolute temperature, $s$: entropy),
The second term represents the kinetic free energy including
the macroscopic motion as well as the mesoscopic mean thermal motion. Notably, the mesoscopic mean thermal motion has the same mathematical form as the macroscopic kinetic energy.
The last part, $\int_\Gamma e_w \delta\Gamma$
is the wall free energy, depicting the wetting effect;
its discussion has been presented elsewhere~\cite{wang2024wetting,wang2024wettinga} and is omitted here for brevity.

Noteworthily, a Lagrange multiplier $\lambda$
has to be added to ensure the volume occupied by the system, namely, $\lambda\int_V\delta V $\footnote{The temperature can also be considered as the 
Lagrange multiplier for the system entropy, $\tau\int_V s \mathrm d V$.}. This Lagrange term is mostly overlooked in existing phase-field models.
As shown in Refs.~\cite{blokhuis1995young,johnson1959conflicts},
the physical meaning of the Lagrange multiplier $\lambda$ is exactly the pressure.
Therefore, the free energy functional ignoring wall free energy should be modified as  
\begin{align}
 E=  \int_V e\delta V+
 \int_V p \delta V;\ e=f(\tau, \phi, \nabla \phi)+ \rho \mathbf{u}\cdot \mathbf{u}.
\end{align}
The Helmholtz-type free energy density $f(\tau, \phi, \nabla \phi)$ 
depends on the temperature $\tau$, composition $\phi$,
and composition gradient $\nabla \phi$, 
as supported by the van der Waals-Cahn theory~\cite{rowlinson1979translation,cahn1958free,cai2024chemo}.
Differing from existing works, we split the entropy into three parts: Boltzmann mixing entropy $s_\phi$, thermal entropy $s_\tau$, and velocity entropy $s_u$:
\begin{equation}
 s=s_\phi+s_\tau+s_u.
\end{equation}
The evolution of these three entropies 
leads to the mass diffusion, heat diffusion, and velocity diffusion equations, respectively,
as demonstrated by the following calculus. 

The variation of the free-energy functional associated with the work leads to the following relation:
\begin{align}
  \frac{\mathrm d E}{\mathrm d  t}&=\int_V\bigg( \mu_\tau \frac{\mathrm d  \tau }{\mathrm d  t}
  + \sum_i \mu_i \frac{\mathrm d  \phi_i  }{\mathrm d  t} 
  +\boldsymbol \mu_u \cdot \frac{\mathrm d  \mathbf u }{\mathrm d t}  
  \bigg)
  \delta V+ \int_V (e+p) \frac{1}{\delta V}\frac{ \mathrm d (\delta V)}{\mathrm d t } \delta V.
  \label{eq:6}
\end{align}
Inside the first integration, the first,  second, and third terms
depict the heat work, 
the chemical work, and the inertial work, respectively,
where the associated potentials are defined as 
\begin{align}
 \mu_\tau&=\frac{\partial e}{\partial \tau};\\
 \mu_i&=\frac{\partial e}{\partial \phi_i}-\nabla\cdot \frac{\partial e}{\partial \nabla \phi_i};\\
 \boldsymbol \mu_u&=\frac{\partial e}{\partial \mathbf{u}}=\rho \mathbf{u}.
\end{align}

The second integration 
in Eq.~\eqref{eq:6} is due to the change of the volume element
in the transport process.
In the approximation of incompressible fluids,
this term vanishes.

Noting that $\mathrm d(\mathrm \delta V)=0$ in the approximation of incompressible fluids 
and realizing the total mathematical derivative (Appendix A):
\begin{align}
\notag
\mathrm d  E=[\mu_\tau \mathrm d \tau+\tau \mathrm d \mu_\tau&+\sum_i (\mu_i \mathrm d \phi_i+\phi_i \mathrm d \mu_i)
+\boldsymbol\mu_u\cdot \mathrm d\mathbf{u}
+\mathbf{u}\cdot \mathrm d\boldsymbol\mu_u]\delta V+ \mathrm d p \delta V,
\end{align}
the heat,  
chemical, and  inertial works in Eq.~\eqref{eq:6}
must be subject to the modified Gibbs-Duhem equation:
\begin{equation}
 \mathrm{d}p+\sum_i  \phi_i  \mathrm{d} \mu_i+\mathbf{u}\cdot  \mathrm{d} \boldsymbol \mu_u+\tau \mathrm{d} \mu_\tau=0.
 \label{eq:17-8}
\end{equation}
By applying standard integration and using the relation,  $\sum_i  \phi_i  \mathrm{d} \mu_i+\mathbf{u}\cdot  \mathrm{d} \boldsymbol \mu_u+\tau \mathrm{d} \mu_\tau=\mathrm{d}(\sum_i  \phi_i \mu_i+\mathbf{u}\cdot  \boldsymbol \mu_u+\tau \mu_\tau )-\mathrm{d}e$,
we obtain a generalized expression for the pressure as:
\begin{equation}
p-p_{ref}=f-\tau\mu_\tau -\sum_i \phi_i \mu_i- \mathbf{u}\cdot \boldsymbol \mu_u.
\label{eq:18-9}
\end{equation}
Here, $p_{ref}$ is an integration constant.

To derive the Gibbs--Duhem relation, Euler's theorem for homogeneous functions is applied to the free-energy element,
\begin{equation}
    e= \mu_\tau \tau \delta V +\mu_i \phi_i \delta V+ \rho \mathbf{u}\cdot  \mathbf{u} \delta V=\mu_\tau \tau  \delta V+
\mu_i \delta V_i^* + \rho \mathbf{u}\cdot  \mathbf{u} \delta V,
\end{equation}
where $e$ is homogeneous of degree one in the extensive variables. This yields the scaling identity
\begin{equation}
 e(\beta \delta V, \beta \delta V_i^*)=  \mu_\tau \tau  \beta\delta V+
\mu_i \beta\delta V_i^* + \rho \mathbf{u}\cdot  \mathbf{u} \beta\delta V =\beta   e( \delta V,  \delta V_i^*).
\end{equation}

 \section{Results and discussion}
 
\subsection{Kinetic equations  of diffusion}
Applying the dissipation principle~\cite{wang2023thermodynamically,zhang2024multi} to the first, second, and third parts in Eq.~\eqref{eq:6}, 
we obtain the heat,
 mass, and velocity diffusion equations:
\begin{align}
 \frac{\mathrm d \tau}{\mathrm d t}&=\nabla\cdot D_\tau\nabla \mu_\tau,\\
 \frac{\mathrm d \phi_i}{\mathrm d t}&=\nabla\cdot D_{\phi_i}\nabla \mu_i,\\
  \frac{\mathrm d \mathbf{u}}{\mathrm d t}&=
  \nabla\cdot D_u (\nabla \boldsymbol\mu_u+\nabla \boldsymbol\mu_u^T),
   \label{eq:120}
\end{align}
where $D_\tau$ and $D_{\phi_i}$ are the mobilities
associated with the heat and mass diffusivities.
$D_1$ depicts the mobility associated with 
the bulk viscosity 
$\eta$ as~\cite{tanaka2000viscoelastic}
\begin{equation}
 D_u=\eta/\rho^2.
 \end{equation}

The acceleration, $\bold a= \mathrm d \bold u/\mathrm d t$, in Newton's law is conventionally related to the mechanical force, $-\nabla p$.
However, it has been recently proposed that the acceleration is rather originated from entropic force~\cite{verlinde2011origin,bianconi2025gravity,bormashenko2023jeans}.
Our concept in Eq.~\eqref{eq:120}
is thus consistent with the entropic origin 
for the acceleration; the time evolution of the velocity is directly related to the velocity entropy  as (Appendix B)
\begin{equation}
    \boldsymbol{\mu}_u\cdot \frac{\mathrm d \bold u }{\mathrm d t}=-\tau \frac{\mathrm d s_u}{\mathrm d  t}.
\end{equation}
Similarly, the time evolution of the 
temperature and composition 
is associated with the thermal and Boltzmann mixing entropy as 
    \begin{align}
   \mu_i \frac{\mathrm d \phi_i }{\mathrm d t}=-\tau \frac{\mathrm d s_{\phi_i}}{\mathrm d  t},\\
   \mu_\tau \frac{\mathrm d \tau }{\mathrm d t}=-\tau \frac{\mathrm d s_\tau}{\mathrm d  t}.
\end{align}

Mass diffusion and momentum diffusion originate from different microscopic exchange processes. In mass diffusion, entropy production arises from the exchange of the positions of atoms or molecules, resulting in the redistribution of matter. In contrast, velocity diffusion is mediated by the electromagnetic fields that are ubiquitously present between neighboring particles. Entropy is produced through the exchange of particles with the field.

\subsection{The origin of momentum evolution}

Dividing by $\mathrm{d} x_k $, $k=1, 2, \cdots, n$ in the modified Gibbs-Duhem relation, Eq.~\eqref{eq:17-8} and using Eq.~\eqref{eq:10-9}, 
we obtain 
\begin{equation}
    -\frac{\mathrm d p}{\mathrm d x_{k}}+ 
    \sum_i \phi_i\frac{\mathrm d \mu_i}{\mathrm d x_k}+ \tau\frac{\mathrm d \mu_\tau}{\mathrm d x_k}+\frac{\mathrm d \bold r}{\mathrm  d x_k}\cdot \frac{\mathrm d (\rho \bold u) }{\mathrm d t}=0.
\end{equation}
In 1D, we have $\mathrm d \bold r/\mathrm  d x=1$
and the above equation becomes 
\begin{equation}
    -\frac{\mathrm d p}{\mathrm d x}+ 
    \sum_i \phi_i\frac{\mathrm d \mu_i}{\mathrm d x}+ \tau\frac{\mathrm d \mu_\tau}{\mathrm d x}+ \frac{\mathrm d (\rho  u) }{\mathrm d t}=0.
\end{equation}
In $n>1$ dimensions,  
by appropriately choosing $\mathrm dx_k$ so that $\mathrm d \bold r/\mathrm  d x_k=1$ as along as the incompressible condition is fulfilled $\mathrm d(\delta V)=0$, paying attention to the fact that $\mathbf u$ and $\mathbf r$ 
has the same direction, and using Eq.~\eqref{eq:16-9},
we obtain 
\begin{equation}
    -\nabla p+ 
    \sum_i \phi_i \nabla \mu_i+  \tau\nabla \mu_\tau+\frac{\mathrm d (\rho \mathbf u) }{\mathrm d t}=0.
    \label{eq:13}
\end{equation}
The physical meaning for the each terms
in the momentum evolution equation is as follows.
The three terms, $\mathrm d(\rho  \mathbf{u})/\mathrm d t$, $ \sum_i  \phi_i \nabla \mu_i$,
and $\tau \nabla \mu_\tau$ are associated with the velocity, Boltzmann mixing, and heat entropies, respectively.
The pressure term is introduced to enforce volume conservation.
Note that viscosity effect has been implicitly considered 
in the time evolution of $\mathrm d   \mathbf{u}/\mathrm d t$
via Eq.~\eqref{eq:120}.

For the approximation of incompressible fluids that $\mathrm d \rho/\mathrm d t=0$,
Eq.~\eqref{eq:13} replicates  Newton's law
\begin{equation}
 \rho\frac{\mathrm d \mathbf{u}}{\mathrm d t}=-\nabla p,
 \label{eq:14}
\end{equation}
when neglecting the interfacial chemical and thermal effects.

It is worth emphasizing that the momentum evolution equation is mathematically equivalent to the proposed generalized Gibbs–Duhem relation. In this sense, the generalized Gibbs–Duhem framework provides a direct connection between Newtonian mechanics and classical thermodynamics.
Unlike the Navier–Stokes formulation, viscous effects are not introduced through the momentum evolution equation. Instead, they are incorporated in Eq.~\eqref{eq:120} through the velocity entropy, thereby embedding dissipation within the thermodynamic description rather than treating it as an explicit mechanical force.

\subsection{Density evolution equation, speed of sound, Bernoulli's law, and equation of state:}
The momentum evolution equation, Eq.~\eqref{eq:13} 
minus the velocity dissipation equation, Eq.~\eqref{eq:120}  leads to a new evolution  equation
for the density as 
\begin{equation}
\notag
 \mathbf{u}\frac{\mathrm d  \rho }{\mathrm d  t}=-\nabla p- \sum_i  \phi_i \nabla \mu_i-\tau \nabla \mu_\tau- \rho \nabla \cdot  [D_u ( \nabla \boldsymbol{\mu}_u+ \nabla \boldsymbol{\mu}_u^T )].
\end{equation}
The validity of this density evolution equation is demonstrated as follows. 

Under isentropic condition, 
the density evolution equation reduces to
\begin{equation}
 \mathbf{u}\frac{\mathrm d  \rho }{\mathrm d  t}=-\nabla p.
 \label{eq25}
\end{equation}
In 1D,
we replicate the classic definition of sound speed~\cite{landau2013fluid}
\begin{equation}
u=\sqrt{\frac{\mathrm d \mathscr P }{\mathrm d\rho}};\  \mathscr P=-p.
\end{equation}
Here, the negative sign denotes compression.
In the classical Gibbs–Duhem equation, the pressure differential is $-\mathrm{d} p$, reflecting only compression work, $p \mathrm{d} (-|\delta V|)$. In contrast, in the present Gibbs–Duhem equation, the pressure differential is $+\mathrm{d} p$, corresponding to a general work term  $p \mathrm{d} (\delta V)$; compression or expansion is characterized by the time evolution of the volume fraction.

By using the relation $\mathrm{d} (\rho \mathbf{u}^2)=\mathbf{u}^2 \mathrm{d} \rho+ 2\rho \mathbf{u}\cdot \mathrm{d}\mathbf{u}$
and the isentropic condition $\rho \mathbf{u}\cdot\mathrm{d}\mathbf{u}= 0$,
Eq.~\eqref{eq25} is further simplified
as 
\begin{equation}
 \mathrm{d}  (p+\rho \mathbf{u}^2)=0.
\end{equation}
This result is nothing but  Bernoulli's equation.

Next, we demonstrate that the density evolution equation recovers the equation of state for an ideal gas. As discussed above, the velocity field is interpreted as containing both the macroscopic flow velocity and the mean thermal velocity. 
In the absence of macroscopic motion, the velocity $\mathbf{u}$ therefore represents the mean thermal velocity of the molecules. At this mesoscopic length scale, the density is defined consistently with this velocity field.
The corresponding kinetic energy density is then given by
\begin{equation}
    \rho \mathbf{u}^2=R_g\tau/v_m,
\end{equation}
where $R_g$ is the universal gas constant and $v_m$
 depicts the molar volume. 
In the 1D steady state, the density evolution for inviscid fluids reduces to 
\begin{equation}
 \rho u^2\frac{1}{\rho}\frac{\mathrm{d} \rho}{\mathrm{d} x}=-\frac{\mathrm{d} p}{\mathrm{d} x},
 \label{eq:100}
\end{equation}
where a constant of unity: $\frac{\rho}{\rho}=1$ has been added to the left hand side.
Integrating from a reference state 
with density $\rho_0$ and pressure $p_0$
to the state to be solved and using the relation  $\rho \mathbf{u}^2=R_g\tau/v_m$,
we obtain the following equation:
\begin{equation}
 R_g\tau(\ln\rho-\ln \rho_0)=(p_0-p)v_m,
\end{equation}
which is the EOS of ideal gas with a factor, $\ln\rho$~\cite{ledesma2014lattice}.

\subsection{Stokes flow}
For incompressible flow, we have $\mathrm d(\rho \mathbf{u})=\rho \mathrm d \mathbf{u}$ and
the generalized Gibbs-Duhem relation without interfacial and thermal terms reads
\begin{equation}
 \mathrm{d}p+\rho \mathbf{u}\cdot \mathrm{d}\mathbf{u}=0.
\end{equation}
Dividing both sides by $\mathrm{d} x_k$, $k=1, 2, \cdots, n$ and adopting the velocity dissipation equation,
$ \mathrm{d} \mathbf{u} / \mathrm{d}t=D_u \nabla^2 \mathbf{u}=(\eta/\rho)\nabla^2 \mathbf{u}$ [Eq.~\eqref{eq:120}],
we recover Stokes's formulation for fluid dynamics:
\begin{equation}
 -\nabla \mathscr P+\eta \nabla^2\mathbf{u}=0.
\end{equation}

\subsection{Remarks}

The classical continuity equation in the Eulerian framework is derived by applying Gauss's theorem to a control volume through which mass may enter and leave. This derivation accounts for the balance of mass flux across the control surface. However, it should be noted that mass transport is inherently accompanied by entropy production, an effect that is not explicitly considered in the application of Gauss's theorem.

In the density evolution equation, the velocity, which includes the mean thermal motion of the particles, can never vanish when the temperature satisfies $\tau > 0$. Furthermore, the density evolution equation shows that the direction of the velocity is determined by the combined effects of the pressure gradient and the entropy gradient. For an isentropic fluid, where the entropy is spatially uniform, the velocity is aligned with the pressure gradient.

For highly compressible fluids, the second integration term  in Eq.~\eqref{eq:6}
does not vanish. One has to solve an additional evolution equation
for $\mathrm d(\delta V)/\mathrm d t$, which is out of the scope of this work.

\section{Conclusion}
We have proposed a generalized Gibbs–Duhem relation that explicitly incorporates kinetic energy:
\begin{equation}
 \mathrm{d}p+\sum_i  \phi_i  \mathrm{d} \mu_i+\mathbf{u}\cdot  \mathrm{d} \boldsymbol \mu_u+\tau \mathrm{d} \mu_\tau=0.
\end{equation}
Remarkably, this generalized relation is mathematically equivalent to the corresponding momentum evolution equation. As a result, the proposed framework establishes a direct connection between Gibbsian thermodynamics and Newtonian mechanics, providing a unified description of nonequilibrium thermodynamics and mechanical processes.

Based on the modified Gibbs–Duhem relation,
we have  derived a new density evolution equation
at the fluid-fluid interface:
\begin{equation}
\notag
 \mathbf{u}\frac{\mathrm d  \rho }{\mathrm d  t}=-\nabla p- \sum_i  \phi_i \nabla \mu_i-\tau \nabla \mu_\tau- \rho \nabla \cdot  [D_u ( \nabla \boldsymbol{\mu}_u+ \nabla \boldsymbol{\mu}_u^T )].
\end{equation}

This result solves the classic high density ratio problem, especially at the water-gas interface
with a density ratio of about 1000.
The derived density evolution equation
is validated 
by reproducing sound speed,
Bernoulli's law, and equation of state for ideal gas,
in limiting cases.

We anticipate that this work will provide new insights into the development of nonequilibrium thermodynamics and stimulate further advances in fluid dynamics.

\appendix*

\section*{Appendix A: Total derivative method}
The present work can also be derived
by expanding the total derivative: 
\begin{align}
 \label{eq:54}
    &\frac{\mathrm d \phi}{\mathrm d t}=\frac{\partial \phi}{\partial t}+\mathbf{u}\cdot \nabla\phi,\\
        &\frac{\mathrm d \nabla \phi}{\mathrm d t}=\frac{\partial \nabla \phi}{\partial t}+\mathbf{u}\cdot  \nabla\nabla\phi.
        \label{eq:55}
\end{align}
The derivative of the
free energy functional reads
\begin{equation}
    \frac{\mathrm d E }{\mathrm d t}=\int_V \frac{\mathrm d (e+p)}{\mathrm d t } \mathrm d V. 
\end{equation}
The total derivative of the pressure reads
\begin{equation}
    \frac{\mathrm d p}{\mathrm d t}=    \frac{\mathrm d p}{\mathrm d \mathbf r}\cdot \frac{\mathrm d \mathbf r }{\mathrm d  t}=\nabla p\cdot \mathbf{u}.
\end{equation}
By using no-flux boundray condition, integration by parts,
and using Eq.~\eqref{eq:54}-\eqref{eq:55} (see Refs.~\cite{wang2023thermodynamically,zhang2024multi}),
the total derivative of $e$ is derived as 
\begin{equation}
 \mathrm d e= [\mu_\tau  \mathrm d\tau+\tau  \mathrm 
 d\mu_\tau +\phi \mathrm d \mu_\phi+\mu_\phi \mathrm d\phi+\rho \mathbf{u}\cdot \mathrm d\mathbf u+ \mathbf{u}\cdot \mathrm d (\rho \mathbf{u})]\delta V.
\end{equation}
Thus, the total time derivative of the free energy functional reads
\begin{equation}
    \frac{\mathrm d E }{\mathrm d t}=\int_V \{\mu_\tau \frac{\mathrm d  \tau }{\mathrm d  t}+\mu_\phi\frac{\mathrm d \phi}{\mathrm d t}+
    \boldsymbol{\mu_u}\cdot \frac{\mathrm d \mathbf{u}}{\mathrm dt}+[\frac{\mathrm d(\rho \mathbf{u})}{\mathrm d t} +\nabla p+\tau\nabla \mu_\tau+\phi\nabla \mu_\phi]\cdot \mathbf{u} \} \mathrm d V.
\end{equation}
The first three terms are associated with the 
heat, mass, and velocity diffusion.
The last term is nothing but the momentum evolution.

\section*{Appendix B: Compatibility of the velocity entropy with special relativity}
For electromagnetic field with effective mass density $\rho$, the kinetic free energy is expressed as $E_k=\int_V \rho \mathbf{c}\cdot \mathbf{c} \mathrm{d} V$, where $\mathbf{c}$ denotes the transport speed of electromagnetic field, i.e., the light speed.
The velocity entropy evolution
is written as 
\begin{equation}
 \rho \mathbf{c}\cdot \frac{\mathrm{d} \mathbf{c} }{\mathrm d t}=-\tau \frac{\mathrm d  s_c}{\mathrm d t},
\end{equation}
where $s_c$ is the velocity entropy.
Considering a 1D setup, we obtain the following equality in the Lagrange framework:
\begin{equation}
 \frac{\mathrm{d}  c}{\mathrm{d}  t}=0;\ \frac{\mathrm d  s_c}{\mathrm d t}=0,
 \label{eq:61}
\end{equation}
as there is no entropy production for light transport.
The velocity entropy theory also enables us to derive the 
classic Maxwell equations; its comprehensive discussion has been presented elsewhere~\cite{wang2024maxwell}.

\section*{Acknowledgments}
I am grateful to Prof. Dr. B. Nestler, L. Bi, and R. Yin for many valuable discussions.
I am also indebted to Prof. H. Stone (Princeton University) and Prof. A. Wagner (North Dakota State University) for their insightful comments, which significantly improved the clarity of this manuscript.


\begin{thebibliography}{10}

\bibitem{landau2013fluid}
Lev~Davidovich Landau and Evgenii~Mikhailovich Lifshitz.
\newblock {\em Fluid Mechanics: Landau and Lifshitz: Course of Theoretical
  Physics, Volume 6}, volume~6.
\newblock Elsevier, 2013.

\bibitem{sprik2025thermodynamics}
Michiel Sprik.
\newblock Thermodynamics of a compressible lattice gas crystal: Generalized
  gibbs--duhem equation and adsorption.
\newblock {\em The Journal of Chemical Physics}, 163(11), 2025.

\bibitem{yue2004diffuse}
Pengtao Yue, James~J Feng, Chun Liu, and Jie Shen.
\newblock A diffuse-interface method for simulating two-phase flows of complex
  fluids.
\newblock {\em Journal of Fluid Mechanics}, 515:293--317, 2004.

\bibitem{worner2012numerical}
Martin W{\"o}rner.
\newblock Numerical modeling of multiphase flows in microfluidics and micro
  process engineering: a review of methods and applications.
\newblock {\em Microfluidics and nanofluidics}, 12(6):841--886, 2012.

\bibitem{wu2024evolution}
Yanchen Wu, Fei Wang, Sai Zheng, and Britta Nestler.
\newblock Evolution dynamics of thin liquid structures investigated using a
  phase-field model.
\newblock {\em Soft Matter}, 20(7):1523--1542, 2024.

\bibitem{reiplinger2022density}
Benedikt Reiplinger and J{\"u}rgen Brillo.
\newblock Density and excess volume of the liquid ti--v system measured in
  electromagnetic levitation.
\newblock {\em Journal of Materials Science}, 57(16):7954--7964, 2022.

\bibitem{caqueret2023density}
Vincent Caqueret, Kaoutar Berkalou, Jean-Louis Havet, Marie Debacq, and
  St{\'e}phane Vitu.
\newblock Density, excess molar volume and vapor--liquid equilibrium
  measurements at 101.3 kpa for binary mixtures containing ethyl acetate and a
  branched alkane: Experimental data and modeling.
\newblock {\em Liquids}, 3(2):187--202, 2023.

\bibitem{wang2024wetting}
Fei Wang and Britta Nestler.
\newblock Wetting and contact-angle hysteresis: Density asymmetry and van der
  waals force.
\newblock {\em Physical Review Letters}, 132(12):126202, 2024.

\bibitem{wang2024wettinga}
Fei Wang, Haodong Zhang, and Britta Nestler.
\newblock Wetting phenomena: Line tension and gravitational effect.
\newblock {\em Physical Review Letters}, 133(24):246201, 2024.

\bibitem{Note1}
The temperature can also be considered as the Lagrange multiplier for the
  system entropy, $\tau \int_V s \mathrm d V$.

\bibitem{blokhuis1995young}
EM~Blokhuis, Y~Shilkrot, and B~Widom.
\newblock Young's law with gravity.
\newblock {\em Molecular Physics}, 86(4):891--899, 1995.

\bibitem{johnson1959conflicts}
Rulon~E Johnson~Jr.
\newblock Conflicts between gibbsian thermodynamics and recent treatments of
  interfacial energies in solid-liquid-vapor.
\newblock {\em The Journal of Physical Chemistry}, 63(10):1655--1658, 1959.

\bibitem{rowlinson1979translation}
John~S Rowlinson.
\newblock Translation of jd van der waals'“the thermodynamik theory of
  capillarity under the hypothesis of a continuous variation of density”.
\newblock {\em Journal of Statistical Physics}, 20:197--200, 1979.

\bibitem{cahn1958free}
John~W Cahn and John~E Hilliard.
\newblock Free energy of a nonuniform system. i. interfacial free energy.
\newblock {\em The Journal of chemical physics}, 28(2):258--267, 1958.

\bibitem{cai2024chemo}
Yuhan Cai, Fei Wang, Haodong Zhang, and Britta Nestler.
\newblock Chemo-elasto-electro free energy of non-uniform system in the diffuse
  interface context.
\newblock {\em Journal of Physics: Condensed Matter}, 36(49):495702, 2024.

\bibitem{wang2023thermodynamically}
Fei Wang, Haodong Zhang, Yanchen Wu, and Britta Nestler.
\newblock A thermodynamically consistent diffuse interface model for the
  wetting phenomenon of miscible and immiscible ternary fluids.
\newblock {\em Journal of Fluid Mechanics}, 970:A17, 2023.

\bibitem{zhang2024multi}
Haodong Zhang, Fei Wang, and Britta Nestler.
\newblock Multi-component electro-hydro-thermodynamic model with phase-field
  method. i. dielectric.
\newblock {\em Journal of Computational Physics}, page 112907, 2024.

\bibitem{tanaka2000viscoelastic}
Hajime Tanaka.
\newblock Viscoelastic phase separation.
\newblock {\em Journal of Physics: Condensed Matter}, 12(15):R207, 2000.

\bibitem{verlinde2011origin}
Erik Verlinde.
\newblock On the origin of gravity and the laws of newton.
\newblock {\em Journal of High Energy Physics}, 2011(4):1--27, 2011.

\bibitem{bianconi2025gravity}
Ginestra Bianconi.
\newblock Gravity from entropy.
\newblock {\em Physical Review D}, 111(6):066001, 2025.

\bibitem{bormashenko2023jeans}
Edward Bormashenko.
\newblock Jeans instability, jeans entropy, and the entropy origin of gravity.
\newblock {\em World J. Phys.}, 1(02):79--86, 2023.

\bibitem{ledesma2014lattice}
Rodrigo Ledesma-Aguilar, Dominic Vella, and Julia~M Yeomans.
\newblock Lattice-boltzmann simulations of droplet evaporation.
\newblock {\em Soft Matter}, 10(41):8267--8275, 2014.

\bibitem{wang2024maxwell}
Fei Wang.
\newblock Thermodynamic Coupling of Mass and Electromagnetic Fields: Entropic Origin of Parity Asymmetry and the Meissner Effect.
\newblock {\em arXiv:2411.16798}.
\end{thebibliography}
\end{document}